\documentclass[prb,twocolumn,showpacs,amsmath,amssymb]{revtex4}

\usepackage{graphicx}
\usepackage{latexsym}
\usepackage{amsmath}
\usepackage{amssymb}
\usepackage{amsfonts}
\usepackage{color}

\newcommand{\ds}{\displaystyle}

\begin{document}

\title{Dephasing time and  magnetoresistance of two-dimensional electron gas
in spatially modulated magnetic fields }

\author{A.~S.~Mel'nikov}
\author{S.~V.~Mironov}
\author{S.~V.~Sharov}
\affiliation{Institute for Physics of Microstructures, Russian
Academy of Science, 603950, Nizhniy Novgorod, GSP-105, Russia}

\date{\today}

\begin{abstract}

The effect of a spatially modulated magnetic field on the weak
localization phenomenon in two-dimensional electron gas (2DEG) is
studied. Both the dephasing time $\tau_H$ and magnetoresistance
are shown to reveal a nontrivial behavior as functions of the
characteristics of magnetic field profiles. The  magnetic field
profiles with rather small spatial scales $d$ and modulation
amplitudes $H_0$ such that $H_0d^2\ll\hbar c/e$
 are characterized by the dephasing rate $\tau_H^{-1}\propto H_0^2d^2$.
The increase in the flux value $H_0d^2$ results in a crossover to
a standard linear dependence $\tau_H^{-1}\propto H_0$. Applying an
external homogeneous magnetic field $H$ one can vary the local
dephasing time in the system and affect the resulting average
transport characteristics. We have investigated the dependence of
the average resistance vs the field $H$ for some generic systems
and predict a possibility to observe a positive magnetoresistance
at not too large $H$ values. The resulting dependence of the
resistance vs $H$ should reveal a peak at the field values $H\sim
H_0$.
\end{abstract}


\maketitle

\section{Introduction}\label{Intro}

The possibility to govern the electronic transport by applying an
inhomogeneous magnetic field has recently attracted considerable
interest. In particular, this problem is intensively studied in
the hybrid ferromagnet/superconductor structures where the
inhomogeneous magnetic field induced by the domain structure in
the ferromagnet or a magnetic dot array is used to control the
superconducting order parameter structure and the transport of
Cooper pairs (see, e.g., Ref.~\onlinecite{buzdin,alad} for
review). It is important to note that the typical values of the
fields used in such experiments can be relatively small: $H\sim
10-10^3 Oe$. Nevertheless in the vicinity of the superconducting
transition even these field values allow to destroy the Cooper
pairs and, thus, strongly affect the electronic transport.

Another possibility to change the conductance applying relatively
weak magnetic fields can be realized even in the normal (i.e.,
nonsuperconducting) structures provided we consider the systems
with measurable quantum interference effects, e.g., disordered
two-dimensional electron gas (2DEG) at low temperatures $T$. In
the latter case the electron conductance is known to be affected
by the weak--localization effects, which are caused by the quantum
interference between the electronic waves propagating along
different time-reversed quasiclassical trajectories \cite{schmid}.
The weak-localization correction $\Delta g $ to the Drude
conductance ${g}_{D} $ in the diffusive limit can be written in
the form
\begin{equation}
\label{eq_1} \Delta g ({\bf r})=-\frac{2e^{2} }{\pi \hbar } D\int
\limits _{\tau }^{\infty }W({\bf r},t_{0} )dt_{0} \ .
\end{equation}
Here $W({\bf r},t_{0})dt_{0}$ is the  probability of electron
return to the point ${\bf r}$ during the time interval
$t_{0}<t<t_0+dt_0$, $\tau$ is the elastic scattering time, $D$ is
the diffusion constant.

In the presence of an external magnetic field the probability of
return is determined by the Green function $C({\bf r}_{f} ,t_{f}
,{\bf r}_{i} ,t_{i} )$ satisfying the so--called Cooperon
equation:
\begin{equation}
\label{eq_2} W({\bf r},t_{0} )=C({\bf r},t_{i} +t_{0} ,{\bf
r},t_{i} ) \ ,
\end{equation}
\begin{equation}
\label{eq_3}
\begin{array}{c}
{\ds \left[\frac{\partial }{\partial {\it t}_{{\it f}} }
+D\left(-i\frac{\partial }{\partial {\bf r}_{f} } -\frac{2e}{\hbar
c} {\bf A}({\bf r}_{f} )\right)^{2} +\frac{1}{\tau _{\varphi } }
\right]C=} \\ {\ds = \delta (t_{f} -t_{i} )\delta ({\bf r}_{f}
-{\bf r}_{i} )\, \ ,}
\end{array}
\end{equation}
where ${\bf A}({\bf r})$ is the vector potential, and
$\tau_{\varphi }$ is the characteristic dephasing time. In the
limit of zero magnetic field the expression for the
weak-localization correction to the conductance obtained from the
Eq.~(\ref{eq_3}) takes the form:
\begin{equation}
\label{eq_4} {\ds \Delta g(H=0)=-\frac{e^{2} }{2\pi ^{2} \hbar }
\ln \frac{\tau _{\varphi } }{\tau } .}
\end{equation}
The maximal size of closed trajectories contributing to this value
is defined by the characteristic dephasing length
 $L_{\varphi} =\sqrt{D\tau _{\varphi } }$.
Applying an external magnetic field perpendicular to the plane of
2DEG system one destroys the coherence for closed trajectories
which enclose the magnetic flux larger than the flux quantum
$\Phi_0=\pi\hbar c /|e|$. The resulting dephasing time $\tau_H$
becomes field dependent and can be obtained by comparing the flux
through the contour of the size $\sqrt{D\tau_H}$ with $\Phi_0$:
$\tau_H^{-1}\sim DH/\Phi_0$. As a consequence, the 2DEG system has
a negative magnetoresistance (see Ref.~\onlinecite{schmid} and
references therein) and the conductance takes the form
\begin{equation}
\label{magn-cond}
\begin{array}{c}
 {\ds \Delta g=-\frac{e^{2} }{2\pi ^{2} \hbar
 }~~~~~~~~~~~~~~~~~~~~~~~~~~~~~~~~~~~~~~~~~~~~~}\\{\ds \times
\left\{\psi \left(\frac{1}{2} +\frac{\hbar c}{4eHD\tau }
\right)-\psi \left(\frac{1}{2} +\frac{\hbar c}{4eHD\tau _{\varphi
} } \right)\right\} \ ,}
\end{array}
\end{equation}
 where $\psi $ is the digamma function. In the
low field limit ($\tau_H\gg\tau_\varphi$) the expression
(\ref{magn-cond}) transforms into the expansion
\begin{equation}
\label{expand} {\ds \Delta g=-\frac{e^{2} }{2\pi ^{2} \hbar }
\left\{\ln \frac{\tau _{\varphi } }{\tau } -\frac{2}{3}
\left(\frac{eHD\tau _{\varphi } }{\hbar c} \right)^{2}
+...\right\} \ .}
\end{equation}

Considering the magnetic fields which are modulated on microscopic
length scales one should modify the above expressions taking into
account the changes in the magnetic flux enclosed by the
interfering trajectories passing through the regions with a
rapidly changing magnetic field.
 The hybrid structures containing the 2DEG systems
 and certain sources of the spatially
modulated magnetic fields attracted recently both the experimental
and theoretical interest
\cite{rammer,bending1,bending2,bending3,bending4,wang,gusev,mathur,ye,xue,ibrahim,shelankov,mancoff}.
 In part, these investigations have been stimulated
 by the possible potential of such systems for making detailed studies
of the inhomogeneous magnetic field distributions.
The magnetic field profiles with microscopic spatial
scales can be induced, e.g., by a
vortex lattice in a superconducting film
 \cite{rammer,bending1,bending2,bending3,bending4},
as well as by a ferromagnetic film domain structure or a magnetic
dot array positioned in the vicinity of the 2DEG system. Note also
that the problem of the 2DEG conductance in a modulated magnetic
field appears to be equivalent to the one of a rough 2DEG layer
placed in a parallel magnetic field \cite{gusev,mathur}. For the
particular case of vortices trapped in a superconducting film the
magnetic field takes the form of flux tubes. An appropriate
theoretical description of the weak--localization phenomenon for
different flux tube radii as compared to the $L_\varphi$ length
has been developed in Ref.~\onlinecite{shelankov}. The
corresponding contribution to the magnetoconductance at low
average fields $H$ appeared to be proportional to the vortex
concentration, i.e. to the $|H|$ value, in contrast to the $H^2$
behavior in a uniform magnetic field. The numerical analysis of
the conductance corrections for the case of a lattice of magnetic
flux tubes for arbitrary relations between the tube radius and
$L_{\varphi }$ was performed in Ref.~\onlinecite{bending4}.
Experimentally these predictions have been confirmed in
Refs.~\onlinecite{bending1,bending2} for GaAs/AlGaAs
heterostructures.

One can expect that the standard expressions (\ref{magn-cond}) and
(\ref{expand}) for local conductance should hold even for the
spatially modulated magnetic fields provided the characteristic
spatial scale $d$ of such modulation is much larger than the size
of the closed trajectories contributing to the conductance
corrections. For a rather strong value of the z-component of the
local field $B({\bf r})$ the latter size can be defined as a
minimum of two lengths: (i) the dephasing length in the absence of
the field $L_\varphi = \sqrt{D\tau_\varphi}$ and (ii) the length
$L_B=\sqrt{D\tau_{B}}=\sqrt{\hbar c/eB({\bf r})}$
 which formally coincides with the textbook definition
of the magnetic length. Here we denote the magnetic field
component along the direction perpendicular to the plane of a 2DEG
system as $B({\bf r})= H+ \delta H({\bf r})$, where $H$ is the
average field value. Thus, considering rather strong fields and/or
not very low temperatures one can use the above expressions for
local conductance substituting the function $B({\bf r})$ instead
of the homogeneous field $H$. This adiabatic picture obviously
breaks down  when the closed interfering trajectories pass through
the regions with rapidly changing magnetic field which happens
either near the zeros of magnetic field or in the limit $d\lesssim
{\rm min}[L_\varphi,L_B]$. The dephasing length and time in this
case are no longer determined by the local field value and their
dependence on the field modulation amplitude $H_0$ can become
rather unusual. In particular, for the magnetic fields with zero
spatial average the dephasing time is proportional to the square
of the field amplitude ($\tau_B^{-1}\propto H_0^2$) which is in
sharp contrast to the linear in $H$ behavior of the dephasing rate
for homogeneous fields. For some model one--dimensional field
profiles such unusual field dependence of the dephasing rate has
been previously predicted in Ref.~\onlinecite{wang}.
Experimentally this behavior $\tau_B^{-1}\propto H_0^2$ has been
observed in Ref.~\onlinecite{gusev} for random magnetic field
profiles.

One of the goals of the present work is to suggest an analytical
description of the weak localization phenomenon in inhomogeneous
magnetic field for a wide class of the field profiles. In Section
II we consider different regimes of the weak localization which
are realized in different regions of magnetic field parameters.
Also in this section we demonstrate that in strong magnetic fields
and/or at not very low temperatures the local approximation is
applicable for calculation of the quantum correction to the
conductance. In Section III we consider the regimes corresponding
to the weak amplitude of magnetic field. In particular, in
subsection \ref{weak_field} we present the calculations of a
natural measurable quantity, i.e., the conductance averaged over
the system area. As a next step, we proceed with the description
of the dephasing rate behavior vs characteristics of the modulated
magnetic field for a wide class of the periodic field profiles
(see Section \ref{def_time}). In Section IV we consider the case
of strong magnetic fields and show that the dependence of the
magnetoresistance vs the average field value appears to reveal an
unusual peak structure. An obvious reason for the non--monotonous
behavior of the resistance vs the average field is associated with
partial field compensation effect which occurs in the regions
where the $z$-components of the average and local fields have the
opposite signs. Thus, applying the external magnetic field to the
system placed in a modulated field with zero average one can
stimulate the interference effects in some regions of the sample.
Depending on the particular shape of the field profile this effect
can result in the negative or positive magnetoresistance of the
sample. In other words, the 2DEG samples coupled with the
subsystems inducing the inhomogeneous magnetic field can reveal a
so--called ``anti--localization'' (see, e.g.,
Refs.~\onlinecite{hikami,bergman}) phenomenon when we apply an
external magnetic field $H$. The results and suggestions for
possible experiments are summarized in Section V.

Hereafter we focus on the case of classically weak magnetic fields
($eB({\bf r})\tau/mc\ll 1$). Indeed the diffusive approximation
for the electron motion is applicable when $B\ll
{\Phi}_{0}/l^{2}$. At such fields the cyclotron frequency
${\omega}_{c}=eB({\bf r})/mc\ll
\left(\hbar/{\varepsilon}_{F}\tau\right)1/\tau \ll 1/\tau$
(${\varepsilon}_{F}$ is the Fermi energy). Thus, these magnetic
fields affect only the interference corrections to the transport
characteristics, and we disregard the inhomogeneous field effect
on the Drude--type contribution to the conductance which has been
previously studied in
Refs.\onlinecite{hedegard,peeters,matulis,geim} on the basis of
the semiclassical approach.

Note that all results obtained in this paper are valid not only
for ideal 2DEG with zero thickness but also for
quasi-two-dimensional electron systems with finite thickness $a$,
which has to satisfy the condition $a\ll L_{\varphi}$. In this
case one can define the field range, in which longitudinal
components do not affect the weak-localization correction to the
conductance, while the transverse component does. Indeed, the
effect of a weak longitudinal component of magnetic field
$H_{\parallel}$ can be described by the renormalization of the
characteristic dephasing time: the value ${\tau}_{\varphi}^{-1}$
has to be replaced by
${\tau}_{\varphi}^{-1}+{\tau}_{H_{\parallel}}^{-1}$ (see, e.g.,
Ref.~\onlinecite{schmid}), where
\[\frac{1}{{\tau}_{H_{\parallel}}}=\frac{1}{3}
{\left(\frac{eH_{\parallel}a}{\hbar c}\right)}^{2}D.\] Thus, the
influence of the  longitudinal component becomes noticeable only
for $H_{\parallel}\sim H_{\parallel}^{*} \sim
{\Phi}_{0}/aL_{\varphi}$. This value is much larger than the
characteristic value $H_{\bot}^{*}\sim {\Phi}_{0}/L_{\varphi}^{2}$
of transverse component, which can strongly affect the
weak-localization correction. Hereafter we assume
$H_{\parallel}\ll H_{\parallel}^{*}$ and neglect
 the effect of longitudinal magnetic field  components.

\section{Different regimes of weak localization and possible
approximate approaches}

\begin{figure}[t!]
\includegraphics[width=0.48\textwidth]{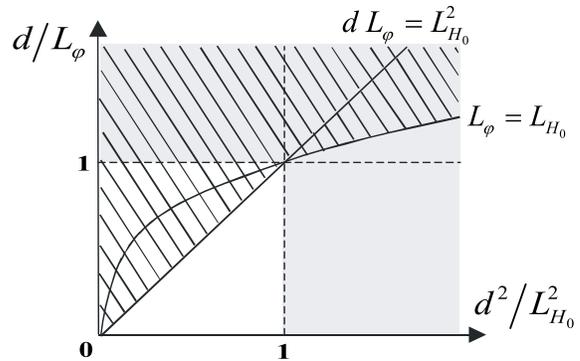}
\caption{The diagram of different weak-localization regimes in the
plane of key parameters. In the gray region the scale of the
magnetic field inhomogeneity $d$ is large enough so that the
dephasing is controlled  by the local magnetic field. In the white
square the magnetic field is weak but its inhomogeneity reveals in
the renormalization of the effective electron dephasing time. In
the shaded region the dephasing occurs at the dephasing length
$L_{\varphi}$ and the influence of magnetic filed reveals in a
small additional correction to the conductivity.}
\label{Fig_diagram}
\end{figure}

In this section we outline the approximate approaches which are
used for describing dephasing regimes in different regions of the
system's parameters.

In the presence of an inhomogeneous magnetic field the
weak-localization correction to the conductance of 2DEG is
determined by the interplay of three lengthscales: (i) dephasing
length $L_{\varphi }$, which at low temperatures grows as
$T^{-1/2}$ (see Ref.~\onlinecite{Deph_Time_T}), (ii) the scale of
the magnetic field inhomogeneity $d$, and (iii) the magnetic
length $L_{H_{0}}=\sqrt{{\Phi}_{0}/H_{0}}$, where $H_{0}$ is the
amplitude of the periodic magnetic field. The ratios of these
lengths define the behaviour of the weak-localization correction
to the conductance. For the analysis of the behavior of quantum
correction to the conductance it is convenient to use the diagram
shown in Fig.~\ref{Fig_diagram}. We choose the parameter
$d/L_{\varphi}$ to describe the temperature dependence of
weak-localization correction to the conductance and the parameter
$d^{2}/L_{H_{0}}^{2}$ to consider the influence of modulated
magnetic field. Note that for a two-dimensional lattice with
translational vectors ${\bf{R}}_{n}=n_{1}d_{1}{\bf
a}_{1}+n_{2}d_{2}{\bf a}_{2}$ (${\bf a}_{1}$, ${\bf a}_{2}$ are
unit vectors and $n_{1}$, $n_{2}$ are integers) the value $d$ is
the absolute value of the smallest vector ${\bf{R}}_{n}$ ($d={\rm
min}\left[d_{1}, d_{2}\right]$).

Depending on the ratio $d/L_{H_0}$ there exist two qualitatively
different mechanisms of the electron dephasing caused by the
inhomogeneous field. Provided $d/L_{H_0}<1$ and simultaneously the
$d$ length scale is smaller than $L_{\varphi}$ (white square in
Fig.~\ref{Fig_diagram}) the dephasing scenario in a modulated
field with zero average can be explained by the following
qualitative arguments. Let us consider a certain
 quasiclassical trajectory of the length $L=\sqrt{D\tau_B}$
  which encloses many primitive cells of
 the periodic field profile. The magnetic flux coming from the
 cells which are positioned inside the contour appears to be
 averaged to zero. The only residual flux is associated with the
 cells which are crossed by the quasiclassical trajectory and give
 a flux contribution which strongly fluctuates with the increase
 in the  area enclosed by the trajectory. The characteristic
 amplitude of these flux fluctuations can be estimated as the
 number of the elementary cells crossed by the trajectory ($L/d$)
multiplied by the typical flux value $H_0d^2$: $\delta\Phi\sim L
dH_0$. Comparing this fluctuating flux with the flux quantum
$\Phi_0$ we find the length of the dephasing $L\sim
\Phi_0/dH_0\sim L_{H_0}^2/d$ and corresponding dephasing rate
$\tau_B^{-1}\sim D d^2H_0^2/\Phi_0^2$. These qualitative arguments
are in beautiful agreement with the quantitative consideration in
section \ref{def_time} carried out on the basis of the ``nearly
free electron'' approximation. In the opposite limit $d/L_{H_0}>1$
the dephasing is controlled by the local magnetic field value.

Of course, the magnetic field provides a dominating dephasing
mechanism only at low temperatures. For rather high temperatures
when $L_{H_{0}}>{\rm
max}\left[L_{\varphi},\sqrt{dL_{\varphi}}\right]$ (shaded region
in Fig.~\ref{Fig_diagram}) the dephasing occurs at the length
$L_{\varphi}$ and one can analyze the magnetic field effect on
 the weak-localization correction to the
conductance perturbatively (see section \ref{weak_field}).

The gray region in Fig.~\ref{Fig_diagram} ($d\gg L_{H_{0}}$ or
 $d>L_{\varphi}$) corresponds to another important
regime: the weak-localization correction to the conductance in
this case can be obtained within the local approximation. This
means that the conductivity at each point of the sample depends on
the local magnetic field. The validity of the local approximation
in this regime can be shown directly from the Eq.~(\ref{eq_3}).
Let us introduce the vectors

\[{\bf R}=\frac{{\bf r}_{f}+{\bf r}_{i}}{2}, ~~~~~~~{\bf r}={\bf
r}_{f}-{\bf r}_{i}.\]

An electron is dephased at the lengthscale which is the minimum of
the scales $L_{\varphi}$ and $L_{H_{0}}$, i.e. only the region
$|{\bf r}|<{\rm min}\left[L_{\varphi},L_{H_{0}}\right]$ makes the
contribution to the weak-localization correction to the
conductance. Therefore in the limit $d\gg {\rm
min}\left[L_{\varphi},L_{H_{0}}\right]$ it is necessary to find
the solution of Eq.~(\ref{eq_3}) only in the case when $|{\bf
r}|\ll d$. In this case we can expand the vector potential ${\bf
A}\left({\bf r}_{f}\right)$:

\[{\bf A}\left({\bf r}_{f}\right)={\bf A}\left({\bf R}+\frac{{\bf
r}}{2}\right)\approx {\bf A}\left({\bf
R}\right)+\frac{1}{2}\left({\bf r},\frac{\partial}{\partial {\bf
R}}\right){\bf A}\left({\bf R}\right).\] Then after introducing a
modified Green function

\[\tilde{C}\left({\bf R},{\bf r}\right)=C\left({\bf R},{\bf
r}\right) {\rm exp} \left[-\frac{2ie}{\hbar c}\left({\bf
A}\left({\bf R}\right),{\bf r}\right)\right]\] one can obtain the
following equation :

\begin{equation}
\label{mod_eq_3_full}
\begin{array}{c}
{\ds \left[\frac{\partial }{\partial {\it t}_{{\it f}} }
+D\left(-i\frac{\partial }{\partial {\bf
r}}-\frac{i}{2}\frac{\partial }{\partial {\bf R}} -\frac{2e}{\hbar
c} \tilde{\bf A}\left({\bf R}, {\bf r}\right)
\right)^{2} +\frac{1}{\tau _{\varphi } } \right]\tilde{C}=} \\
{\ds ~~~~~~~~~~~~~~~= {\rm exp} \left[-\frac{2ie}{\hbar
c}\left({\bf A}\left({\bf R}\right),{\bf r}\right)\right]\delta
({\bf r})}\delta (t_{f}-t_{i} ),
\end{array}
\end{equation}
where

\[\tilde{\bf A}\left({\bf R}, {\bf
r}\right)=\frac{1}{2}\left[\tilde{\bf H}\left({\bf R}\right),{\bf
r}\right], ~~~~~~\tilde{\bf H}\left({\bf
R}\right)=\left[\frac{\partial }{\partial {\bf R}},{\bf
A}\left({\bf R}\right)\right].\] The right part of
Eq.~(\ref{mod_eq_3_full}) contains $\delta ({\bf r})$, so we can
put ${\bf r}=0$ in the exponential prefactor.

Note, that in Eq.~(\ref{mod_eq_3_full}) one can neglect the term
containing the derivative $\partial/\partial {\bf R}$. Indeed, it
has the order $d^{-1}$ whereas the value $\tilde{\bf A}\left({\bf
R}, {\bf r}\right)$ and the term containing the derivative
$\partial/\partial {\bf r}$ have the order $1/{\rm
min}\left[L_{\varphi},L_{H_{0}}\right]$, so the terms containing
$\partial/\partial {\bf R}$ are negligible. In this case the
Eq.~(\ref{mod_eq_3_full}) takes the form

\begin{equation}
\label{mod_eq_3}
\begin{array}{c}
{\ds \left[\frac{\partial }{\partial {\it t}_{{\it f}} }
+D\left(-i\frac{\partial }{\partial {\bf r}} -\frac{2e}{\hbar c}
\tilde{\bf A}\left({\bf R}, {\bf r}\right)
\right)^{2} +\frac{1}{\tau _{\varphi } } \right]\tilde{C}=} \\
{\ds ~~~~~~~~~~~~~~~~~~~~~~~~~~~~~~~~= \delta (t_{f}-t_{i} )\delta
({\bf r})}.
\end{array}
\end{equation}

The equation (\ref{mod_eq_3})  formally coincides with the
Eq.~(\ref{eq_3}) for the case of homogeneous magnetic field
$\tilde{\bf H}\left({\bf R}\right)$, which depends on the variable
${\bf R}$ as a parameter. Thus, in the gray region in the
Fig.~\ref{Fig_diagram} one can use the local approximation to
calculate the weak-localization correction to the conductance. In
what follows we show that in this case the spatially modulated
magnetic field with zero average can cause the effect of positive
magnetoresistance of 2DEG in the external homogeneous magnetic
field.

Note, that the local approximation breaks down at the points where
the magnetic field is changing rapidly, i.e. near the zeros of the
magnetic field. Nevertheless considering the spatially averaged
conductance one can neglect the correction coming from these
regions which appears to be small in the limit $L_{\varphi}/d\ll
1$.

\section{Weak magnetic fields. Negative magnetoresistance}

\subsection{Magnetoresistance of 2DEG. Second order perturbation theory.}
\label{weak_field}

\subsubsection{Quantum correction to the conductance in the field
with arbitrary spatial configuration}\label{Sec_2_2}

Let us consider the case of magnetic field with arbitrary spatial
configuration but with zero spatial average. In this subsection we
find an analytical solution of the equation (\ref{eq_3}) in the
extreme case of low magnetic field. This means that the magnetic
flux through any closed contour of the size
$\sqrt{D\tau_{\varphi}}$ is mush less than $\Phi _{0}$. In this
case the magnetic field weakly affects the weak-localization
correction to the conductance, and the equation (\ref{eq_3}) can
be solved within the frames of the perturbation theory with a
small parameter proportional to the value of magnetic field.

Let us introduce the Fourier transform of the magnetic field:

\begin{equation}
\label{Two_dim_field} H_{z}({\bf r})=\int \limits _{-\infty
}^{\infty }\int \limits _{-\infty }^{\infty }H_{{\bf k}}e^{i{\bf
k}{\bf r}}d^{2}{\bf k}.
\end{equation}
Here ${\bf r}$ is a vector in the plane of 2DEG. We assume all
spatial harmonics $H_{{\bf k}}$ of magnetic field to be small.

The corresponding vector potential can be chosen in the form
\begin{equation}
\label{Two_dim_potential} {\bf A}({\bf r})=\int \limits _{-\infty
}^{\infty }\int \limits _{-\infty }^{\infty } {\bf A}_{{\bf
k}}e^{i{\bf k}{\bf r}}d^{2}{\bf k},
\end{equation}
where
\begin{equation}
\label{A_k} {\bf A}_{{\bf k}}=\frac{i\left[{\bf k},{\bf
z}_{0}\right]}{k^{2}}H_{{\bf k}}.
\end{equation}

In the zero order of the small parameter we considered the Green
function as the one without magnetic field:

\[C_{0} =\frac{1}{4\pi Dt_{0}} \exp
\left(-\frac{r_{0}^{2}}{4Dt_{0}} -\frac{t_{0}}{\tau _{\varphi }}
\right).\] Here ${\bf r}_{0} ={\bf r}_{f} -{\bf r}_{i} $, $t_{0}
=t_{f} -t_{i} $. Further we represent the operator in the left
part of the Eq. (\ref{eq_3}) as a sum of operators
$\hat{F}=\hat{F}_{0} +\hat{F}_{1} +\hat{F}_{2} $ where

\[\hat{F}_{0} =\left[\frac{\partial }{\partial {\it t}_{{\it f}} }
-D\frac{\partial ^{2} }{\partial {\bf r}_{f}^{~2} } +\frac{1}{\tau
_{\varphi } } \right],\]

\[\ds \hat{F}_{1} =\frac{4eD}{\hbar c}\int \limits _{-\infty
}^{\infty }\int \limits _{-\infty }^{\infty }d^{2}{\bf k}~H_{{\bf
k}}~e^{i{\bf k}{\bf r}_{f}}\left[{\bf k},\frac{\partial }{\partial
{\bf r}_{f} }\right],\]

\[\hat{F}_{2} =\frac{4e^{2} D}{\hbar ^{2} c^{2}}
{\left[\int \limits _{-\infty }^{\infty }\int \limits _{-\infty
}^{\infty }d^{2}{\bf k}~ \frac{i\left[{\bf k},{\bf
z}_{0}\right]}{k^{2}}H_{{\bf k}} e^{i{\bf k}{\bf r}}\right]}^{2}
.\]

In the first order of perturbation theory the correction to the
Green function can be written as

\[\begin{array}{c} C_{1} ({\bf r}_{f} ,t_{f} ,{\bf r}_{i} ,t_{i}
,) =\\ =-\int \limits _{t_{i} }^{t_{f} }\int \limits _{-\infty
}^{+\infty }\int \limits _{-\infty }^{+\infty }C_{0} ({\bf r}_{f}
,t_{f} ,{\bf r}',t')\hat{F}_{1} C_{0} ({\bf r}',t',{\bf r}_{i}
,t_{i} )d{\bf r}'dt'   .\end{array}\] If $(x_{f} ,y_{f} )\to
(x_{i} ,y_{i} )$ then $C_{1} =0$. This reflects the fact that the
quantum correction does not depend on the sign of applied magnetic
field. The second order correction to the Green function is
defined by the expression

\begin{equation}
\label{eq_7} \begin{array}{c} C_{2} =-\int \limits _{t_{i}
}^{t_{f} }\int \limits _{-\infty }^{+\infty }\int \limits
_{-\infty }^{+\infty }C_{0} ({\bf r}_{f} ,t_{f} ,{\bf r}',t')
\\ \times \left[\hat{F}_{1} C_{1} ({\bf r}',t',{\bf r}_{i} ,t_{i}
)+\hat{F}_{2} C_{0} ({\bf r}',t',{\bf r}_{i} ,t_{i} )\right]d{\bf
r}'dt'   .\end{array}
\end{equation}
Let us introduce the value $\Delta {g}_{H}$

\begin{equation}
\label{delta} \Delta {g}_{H} =\Delta g({\bf B})-\Delta g(0),
\end{equation}
where the $\Delta g({\bf B})$ is the weak-localization correction
to the conductance in the inhomogeneous magnetic field ${\bf
B}({\bf r})$. Then the value $\Delta {g}_{H}$ is determined by the
second order correction $C_{2}({\bf r},t_{0})$:

\begin{equation}
\label{g2} \Delta {g}_{H}({\bf r})=-\frac{2e^{2}D }{\pi \hbar }
\int \limits _{0}^{\infty }C_{2}({\bf r},t_{0} )dt_{0}.
\end{equation}
In the expression (\ref{g2}) we put the lower integration limit
equal to zero because of the absence of the small $t_{0}$
divergence in the integrand. Thus, we neglect the correction of
the order $\tau / {\tau}_{\varphi}$.

As we are interested only in spatially averaged correction to the
conductance the expression (\ref{g2}) should be integrated over
${\bf r}$. Performing the integration in the expression
(\ref{eq_7}) we obtain the averaged correction to the conductance
$\left\langle \Delta {g}_{H} \right\rangle $:

\begin{equation}
\label{Two_dim_res} \begin{array}{c} {\ds \left\langle \Delta
{g}_{H} \right\rangle =\frac{8e^{4} D}{\hbar ^{3} c^{2} S} \int
\limits _{0}^{\infty }dt_{0} \, e^{-\frac{t_{0} }{\tau _{\varphi }
} } \int \limits _{-\infty }^{\infty }\int \limits _{-\infty
}^{\infty }d^{2}{\bf k}\frac{H_{{\bf k}} H_{-{\bf k}} }{k^{2}
}}\\{\ds \times\left(1-\frac{2}{k\sqrt{Dt_{0} } } e^{-\frac{k^{2}
Dt_{0} }{4} } \Phi \left(\frac{k\sqrt{Dt_{0} } }{2} \right)\right)
,}
\end{array}
\end{equation}
where $S$ is the area of the sample, $\Phi (\xi )=\int \limits
_{0}^{\xi }e^{t^{2} } dt $. Since we neglect the corrections
proportional to $\tau / {\tau}_{\varphi}$ the integration in the
expression (\ref{Two_dim_res}) should be performed over $|{\bf
k}|<{(D\tau)}^{-1/2}$.

Note that $H_{-{\bf k}}=H_{{\bf k}}^{*}$ and for $\alpha>1$
\begin{equation}
\int \limits_{0}^{\infty} e^{-\alpha
x^{2}}\Phi\left(x\right)dx=\frac{1}{4\sqrt{\alpha}}
\ln\left(\frac{\sqrt{\alpha}+1}{\sqrt{\alpha}-1}\right).
\end{equation}
Then the expression (\ref{Two_dim_res}) can be rewritten in the
form

\begin{equation}
\label{Two_dim_full_res} \ds \left\langle \Delta {g}_{H}
\right\rangle =\frac{2e^{4} D^{2}\tau _{\varphi }^{2}}{\hbar ^{3}
c^{2} S } \int \limits _{-\infty }^{\infty }\int \limits _{-\infty
}^{\infty }d^{2}{\bf k}~{\left|H_{{\bf k}}\right|}^{2}~
F\left(\frac{k^{2}D\tau _{\varphi }}{4}\right),
\end{equation}
where
\begin{equation}
\label{eq_F} F(z)=\frac{1}{z}\left(1-\frac{1}{\sqrt{z(z+1)}}
\ln\left(\sqrt{z+1}+\sqrt{z}\right)\right).
\end{equation}

\begin{figure}[t!]
\includegraphics[width=0.4\textwidth]{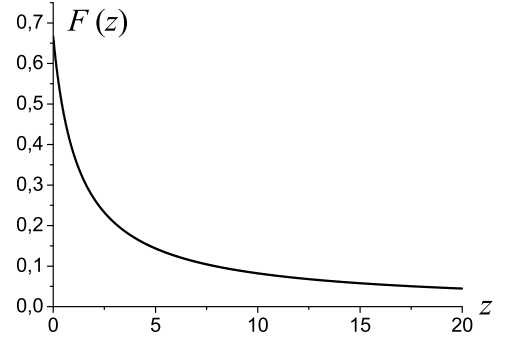}
\caption{The function $F(z)$ defined by the expression
(\ref{eq_F}).} \label{F_z}
\end{figure}

With the increase in the $z$ coordinate the function $F(z)$ is
monotonically decreasing from the value $2/3$ at $z=0$ to zero at
$z=\infty$ decaying as $z^{-1}$ at large $z$ values. Therefore,
the spatial harmonics of magnetic field with $\left|{\bf
k}\right|\ll L_{\varphi }^{-1} $ make the main contribution to the
weak-localization correction. In particular, for the case of
magnetic field with narrow spectrum in the momentum space (the
value $H_{{\bf k}} $ is non-zero only in the spectral region
$\left|{\bf k}\right|\ll L_{\varphi }^{-1} $) the value
$\left\langle \Delta {g}_{H} \right\rangle $ is defined by the
expression
\begin{equation}
\label{Two_dim_small_k} \ds \left\langle \Delta {g}_{H}
\right\rangle =\frac{4e^{4} D^{2}\tau _{\varphi }^{2}}{3\hbar ^{3}
c^{2} S } \int \limits _{-\infty }^{\infty }\int \limits _{-\infty
}^{\infty }~{\left|H_{{\bf k}}\right|}^{2}~d^{2}{\bf k}.
\end{equation}
Using the properties of Fourier transformation, we can rewrite the
expression (\ref{Two_dim_small_k}) in the form
\begin{equation}
\label{Two_dim_small_k_H} \ds \left\langle \Delta {g}_{H}
\right\rangle =\frac{e^{4} D^{2}\tau _{\varphi }^{2}}{3
{\pi}^{2}\hbar ^{3} c^{2} S } \int \limits
_{S_{0}}~H_{z}^{2}d^{2}{\bf r}.
\end{equation}
It is seen from the Eq.~(\ref{Two_dim_small_k_H}) that in the case
of weak non-homogeneous field with spatial scale larger than
$L_{\varphi } $ the averaged correction to conductance is defined
only by the square of magnetic field averaged over the sample.
This result corresponds to the local approximation.

Note that the expression (\ref{Two_dim_res}) is correct also in
the case of magnetic field with non-zero spatial average $H$ which
satisfies the condition $HL_{\varphi}^{2}\ll {\Phi}_{0}$. In this
case the expression for the spatially averaged weak-localization
correction to the conductance has the form

\begin{equation}
\label{eq_comb}\begin{array}{c} {\ds
\left<g(H)\right>=g_{D}-\frac{e^{2} }{2\pi ^{2} \hbar } \ln
\frac{\tau _{\varphi } }{\tau }+\left\langle \Delta {g}_{H}
\right\rangle }\\{\ds +\frac{e^{2} }{3\pi ^{2} \hbar
}\left(\frac{eHD\tau _{\varphi } }{\hbar c} \right)^{2},}
\end{array}
\end{equation}
where $\left\langle \Delta {g}_{H} \right\rangle $ is defined by
the expression (\ref{Two_dim_full_res}) for magnetic field with
zero average. Thus the homogeneous component of the magnetic field
makes small additional contribution to the averaged correction.

The expression (\ref{g2}) for the local conductance value can be
further simplified for the particular case of one-dimensional
field which depends on the $x$ coordinate. In this case the
magnetic field can be written in the form

\[H_{z} =\int \limits _{-\infty }^{\infty }H_{k} e^{ikx} dk ,\]
where $k$ is the scalar Fourier variable. Performing integration
in (\ref{eq_7}) we obtain an analytical expression for the Green
function $C_{2}({\bf r})$:

\begin{equation}\label{eq_arbgreen}\begin{array}{c} {C_{2} =\frac{4e^{2} e^{-\frac{t_{0} }{\tau
_{\varphi } } } }{\pi \hbar ^{2} c^{2} \left(Dt_{0} \right)^{3/2}
} \int \limits _{-\infty }^{\infty }dk\int \limits _{-\infty
}^{\infty }dq\frac{H_{k} H_{q} }{k^{2} q^{2} (k+q)} e^{i(k+q)x}}
\\
{\times \left[ e^{-\frac{(q+k)^{2} Dt_{0} }{4} }\Phi
\left(\frac{\left(k+q\right)\sqrt{Dt_{0} } }{2} \right) \right.}
\\
{-e^{-\frac{Dk^{2} t_{0} }{4} }\Phi\left(\frac{k\sqrt{Dt_{0} }
}{2} \right) -e^{-\frac{q^{2} Dt_{0} }{4} } \Phi
\left(\frac{q\sqrt{Dt_{0} } }{2} \right)}
\\
{\ds \left. +\frac{kqDt_{0} }{2} e^{-\frac{(k+q)^{2} Dt_{0} }{4} }
\Phi \left(\frac{(k+q)\sqrt{Dt_{0} } }{2} \right) \right] .}
\end{array}\end{equation}
This expression can be made even more transparent in the special
case of the magnetic field with the sinusoidal profile.

\subsubsection{Quantum correction to the conductance in
low sinusoidal magnetic field}

Let the magnetic field has the form

\begin{equation}
\label{F_cos} H_{z}\left(x\right)=H_{0}\cos\left(kx\right) \ .
\end{equation}
Then for the value $\Delta {g}_{H} $ we obtain the following
expression:

\begin{equation}
\label{eq_8}
\begin{array}{c} {\ds \Delta {g}_{H} =\frac{2e^{4} H_{0}^{2}
}{\pi ^{2} \hbar ^{3} c^{2} k^{4} } \int \limits _{0}^{\infty
}e^{-\frac{\beta ^{2} }{k^{2} D\tau _{\varphi } } } \left(\beta
-2e^{-\frac{\beta ^{2} }{4} } \Phi \left({\tfrac{\beta }{2}}
\right)\right.  } \\ {\ds  +\left. \cos (2kx)\frac{e^{-\beta ^{2}
} }{\beta ^{2} } \left\{4e^{\frac{3}{4} \beta ^{2} } \Phi
\left({\tfrac{\beta }{2}} \right)-(\beta ^{2} +2)\Phi \left(\beta
\right)\right\}\right)d\beta .} \end{array}
\end{equation}
One can see that the only dimensionless parameter $\alpha
=k\sqrt{D\tau _{\varphi } } =\pi L_{\varphi } /d$ defines the
value of the integral in the expression (\ref{eq_8}).

\begin{figure}[t!]
\includegraphics[width=0.48\textwidth]{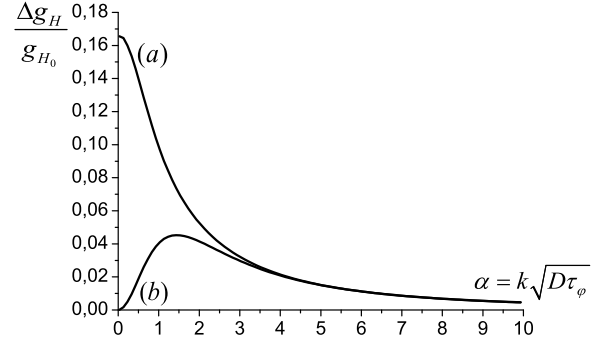}
\caption{The dependence of the value $\Delta {g}_{H} $ vs
parameter $\alpha $ for $x=d/2$ (a) and $x=d/2$ (b).}
\label{Fig_phase_int}
\end{figure}

It is interesting to consider the dependence of the value $\Delta
{g}_{H} $ on the period of magnetic field $d$. The dependencies of
the value $\Delta {g}_{H} $ on the parameter $\alpha  =\pi
L_{\varphi } /d$ for two interesting cases $x=0$ and $x=d/2$ are
shown in Fig.~\ref{Fig_phase_int}, where we introduced the value

\[{g}_{H_{0}}=\frac{2e^{2}}{\hbar}
{\left(\frac{H_{0}D{\tau}_{\varphi}} {{\Phi}_{0}}\right)}^{2}.\]

Note that the value $\Delta {g}_{H}({\bf r})$ is defined by the
module of the averaged magnetic flux through all possible closed
trajectories of the size $L_{\varphi }$ which are passing through
the point ${\bf r}$. In the maxima of the magnetic field ($x=nd$,
$n$ is an integer) the averaged flux is decreasing with $d$
decreasing (increasing $\alpha $) and this leads to the decrease
in the $\Delta {g}_{H}$ value. At the zeroes of the magnetic field
($x=d/2+nd$) when the scales $d$ and $L_{\varphi } $ are
comparable one can observe a maximum in the dependence $\Delta
{g}_{H}$ vs $\alpha $ (see the curve (b) in
Fig.~\ref{Fig_phase_int}). In this case the averaged module of the
magnetic flux through the closed trajectories is maximal.

The analysis of the expression for $\Delta {g}_{H} $ shows that in
the limit of $d\ll L_{\varphi } $ the value $\Delta {g}_{H} $ is
proportional to $d^{2}$:

\[\Delta {g}_{H} \approx \frac{e^{2} }{\pi ^{2} \hbar }
\left(\frac{L_{\varphi } }{L_{H_{0}} } \right)^{4} \cdot
\left(\frac{d}{L_{\varphi } } \right)^{2} ,\] where
$L_{H_{0}}=\sqrt{{\Phi}_{0}/H_{0}}$.

In case of smooth field variation ($d\gg L_{\varphi } $, but
$H_{0} dL_{\varphi } \ll \Phi _{0} $) keeping the corrections
$\propto{\alpha}^{2} $ we find:

\begin{equation}
\label{eq_9} \Delta {g}_{H} =\frac{e^{2} L_{\varphi }^{4} }{6\hbar
\, L_{H_{0}}^{4} } \left((1+\cos 2kx)-\frac{\alpha ^{2} }{5}
(1+7\cos 2kx)\right).
\end{equation}
The expression (\ref{eq_9}) differs strongly from the
correspondent expression for the case of homogeneous field, since
even at the points of zero  magnetic field  the value $\Delta
{g}_{H} $ is positive. This fact is quite natural since the
averaged module of the flux through the closed trajectories does
not vanish even at these points.

\subsection{Dephasing time in spatially periodic magnetic fields}
\label{def_time}

The expressions (\ref{Two_dim_full_res}) and (\ref{eq_comb})
obtained within the perturbation theory diverge at low
temperatures as the dephasing time ${\tau}_{\varphi}$ tends to
infinity. Thus, to describe the behavior of the conductance at low
temperatures one should take account of the renormalization of the
dephasing time caused by the magnetic field.

In this subsection we consider such renormalization procedure for
a periodic magnetic field

\begin{equation}
\label{A_periodic} H_{z}\left({\bf r}+{\bf R}_{n}\right) =
H_{z}\left({\bf r}\right),
\end{equation}
where ${\bf R}_{n}$ are translational vectors which generate a
two-dimensional lattice.

The expression for the electron probability of return obtained
from the solution of the equation (\ref{eq_3}) can be written in
the following form:

\begin{equation}\label{eq_13}\ds
W({\bf r},t_{0} )= e^{-\frac{t_{0} }{\tau _{\varphi } } } \sum
\limits _{j}\left|\psi _{j} ({\bf r})\right|^{2}
e^{-\varepsilon_{j}Dt_{0} } .
\end{equation}
Here $\varepsilon_{j}$ and $\psi _{j} ({\bf r})$ are the
 eigenvalues and normalized eigenfunctions of the operator

\begin{equation}
\label{H_oper} \ds \hat{H}({\bf r})=\left(-i\nabla
-\frac{2e}{\hbar c} {\bf A}({\bf r})\right)^{2},
\end{equation}

\begin{equation}\label{eigen}
\hat{H}({\bf r})\, \psi _{j} ({\bf r})=\varepsilon_{j}\psi _{j}
({\bf r}),
\end{equation}

\[\int \limits _{-\infty }^{+\infty }\int \limits _{-\infty }^{+\infty }\psi _{j} ({\bf r})\psi
_{j'}^{*} ({\bf r})d^{2}{\bf r} =\delta_{j,j'}.\] Substituting the
expression (\ref{eq_13}) into the Eq.~(\ref{eq_1}) we find the
quantum correction to the conductance

\begin{equation}
\label{eq_14} \ds \Delta g\left({\bf r}\right)=-\frac{2e^{2} }{\pi
\, \hbar } \sum \limits _{j}{\frac{\left|\psi _{j} ({\bf
r})\right|^{2} }{\varepsilon_{j}+\frac{1}{D\tau _{\varphi } } }}
 e^{-D\tau \left(\varepsilon_{j}+\frac{1}{D\tau
_{\varphi } } \right)} .
\end{equation}

Thus, the correction to the conductance is defined only by the
spectrum and by the set of eigenfunctions of the operator
$\hat{H}({\bf r})$. Further we will be interested only in
spatially averaged quantum correction to the conductance which
defines the voltage between the sample contacts. Then taking into
account the normalization condition for eigenfunctions we obtain:

\begin{equation}\label{eq_15}\ds
\left\langle \Delta g\right\rangle =-\frac{2e^{2} }{\pi \, \hbar }
\sum \limits _{j}{\frac{\ds e^{-D\tau
\left(\varepsilon_{j}+\frac{1}{D\tau _{\varphi } } \right)}}{
\varepsilon_{j}+\frac{1}{D\tau _{\varphi } } }} .
\end{equation}

To calculate the spectrum of the operator $\hat{H}({\bf r})$ let
us expand the magnetic field into the Fourier series

\begin{equation}
\label{H_Fourier} B_{z}\left({\bf r}\right) = H+\sum \limits_{{\bf
b}_{n} \not= 0} H_{n} e^{i {\bf b}_{n} {\bf r}}.
\end{equation}
We start our analysis from the case of magnetic field with zero
spatial average, i.e. we put $H=0$. The corresponding vector
potential can also be written in the form of Fourier series

\begin{equation}
\label{A_Fourier} {\bf A}\left({\bf r}\right) = \sum \limits_{{\bf
b}_{n} \not= 0} {\bf A}_{n} e^{i {\bf b}_{n} {\bf r}}.
\end{equation}

Choosing the vector potential in the Lorentz gauge div${\bf A}=0$,
so that

\begin{equation}
\label{A_n} {\bf A}_{n}=i\frac{\left[{\bf b}_{n},{\bf H}_{n}
\right]}{{\bf b}_{n}^{2}},
\end{equation}
we obtain the following expression for the operator $\hat{H}({\bf
r})$:

\begin{equation}
\label{eq_operator} \begin{array}{c}{\ds \hat{H}({\bf
r})=-\nabla^{2}-\frac{4e}{\hbar c} \sum \limits _{{\bf b}_{n}\not=
0} H_{n} e^{i{\bf b}_{n} {\bf r}}{\bf \alpha}_{n}\nabla}\\{\ds
+\frac{4e^{2}}{\hbar^{2}c^{2}} \sum \limits _{{\bf b}_{n}\not=
0}Q_{n} e^{i{\bf b}_{n} {\bf
r}}+\frac{4e^{2}}{{\hbar}^{2}c^{2}}\sum \limits _{{\bf b}_{n}\not=
0} \frac{{|H_{n}|}^{2}}{{\bf b}_{n}^{2}}}\end{array},
\end{equation}
where
\begin{equation}
{\bf \alpha}_{n}=\frac{\left[{\bf b}_{n},{\bf z}_{0}\right]}{{\bf
b}_{n}^{2}},
\end{equation}
\begin{equation}
Q_{n}=\sum \limits _{m\not= 0,m \not=n} \frac{H_{m}
H_{n-m}}{{\left({\bf b}_n-{\bf
b}_{m}\right)}^{2}}\left(1-\frac{\left({\bf b}_{n}, {\bf
b}_{m}\right)}{{\bf b}_{m}^{2}}\right).
\end{equation}

The Hamiltonian is translationally invariant ($\hat{H}({\bf
r}+{\bf R}_{n})=\hat{H}({\bf r})$) and  its eigenfunctions satisfy
the {\it Bloch} theorem:
\begin{equation}
\label{psi} \psi_{{\bf k}}({\bf r})=\sum \limits _{{\bf b}_{n}}
u_{n} e^{i({\bf k}+{\bf b}_{n}){\bf r}}.
\end{equation}
Substituting the expression (\ref{psi}) into Eq.~(\ref{eigen}) and
introducing the value
\begin{equation}
\label{eq_newepsilon}
\varepsilon'=\varepsilon-\frac{4e^{2}}{{\hbar}^{2}c^{2}}\sum
\limits _{{\bf b}_{n}\not= 0} \frac{{|H_{n}|}^{2}}{{\bf
b}_{n}^{2}},
\end{equation}
we find the following equation for the amplitudes $u_{n}$:

\begin{equation}
\label{eq_un}
\begin{array}{c}{\ds
\left[{\left({\bf k}+{\bf
b}_{n}\right)}^{2}-\varepsilon'\right]u_{n}+\frac{4e^{2}}{\hbar^{2}c^{2}}\sum
\limits _{{\bf b}_{m}\not= 0} Q_{m}u_{n-m}}\\{\ds-\frac{4ie}{\hbar
c}\sum \limits _{{\bf b}_{m}\not= 0} u_{n-m}H_{m}{\bf
\alpha}_{m}\left({\bf k}+{\bf b}_{n}-{\bf b}_{m}\right)=0.}
\end{array}
\end{equation}
In the absence of the magnetic field only the amplitude $u_{n}$
corresponding to $n=0$ is nonzero ($u_{n=0}\not= 0$ and $u_{n
\not=0}= 0$). Therefore, in this case the spectrum has the form
$\varepsilon({\bf k})={\bf k}^{2}$.

Let us consider the dephasing time at low temperatures when
$L_{\varphi}\gg d$ ($d$ is the characteristic scale of magnetic
field inhomogeneity). We also restrict ourselves to the case of
low amplitude of the magnetic field so that $L_{{H}_{0}}\gg d$
($L_{{H}_{0}}=\sqrt{{\Phi}_{0}/H_{0}}$). Then, as it is seen from
the Eq.~(\ref{eq_15}), the zero temperature divergence of the
conductance correction comes from the region of low $\varepsilon$
which corresponds to the region of low $|{\bf k}|$. Taking into
account the condition $d\ll L_{{H}_{0}} $ we will assume that the
region $|{\bf k}|\ll |{\bf b}_{n}|$ gives the main contribution to
the low temperature correction. In this case the spectrum of the
operator $\hat{H}({\bf r})$ can be calculated in the ``nearly free
electron'' approximation.

In the presence of magnetic field the spectrum can be written in
the form $\varepsilon'({\bf k})={\bf k}^{2}+\varepsilon
^{(1)}({\bf k}) +\varepsilon ^{(2)} ({\bf k})$, where $\varepsilon
^{(1)}$ is proportional to the field amplitude and $\varepsilon
^{(2)}$ is proportional to the square of the field's amplitude.

The first order correction to the zero field spectrum is equal to
the second term in the expression (\ref{eq_newepsilon}) but with
the opposite sign. Thus, the correction $\varepsilon ^{(1)}$ in
the spectrum $\varepsilon'$ is zero. The second-order correction
$\varepsilon ^{(2)} $ has the following form:

\begin{equation}\label{eq_16}
\varepsilon ^{(2)} = -\frac{16e^{2}}{\hbar^{2}c^{2}}\sum
\limits_{{\bf b}_{n}\not=0} \frac{{|H_{n}|}^{2}{\left({\bf
\alpha}_{n},{\bf k}\right)}^{2}}{{\bf b}_{n}^{2}}.
\end{equation}
Thus, the spectrum reads

\begin{equation}\label{eq_17}\begin{array}{c}{\ds
\varepsilon({\bf k})={\bf
k}^{2}-\frac{16e^{2}}{\hbar^{2}c^{2}}\sum \limits_{{\bf
b}_{n}\not=0} \frac{{|H_{n}|}^{2}{\left({\bf k},\left[{\bf
b}_{n},{\bf z}_{0}\right]\right)}^{2}}{{\bf
b}_{n}^{6}}}\\{\ds+\frac{4e^{2}}{{\hbar}^{2}c^{2}}\sum \limits
_{{\bf b}_{n}\not= 0} \frac{{|H_{n}|}^{2}}{{\bf b}_{n}^{2}}.}
\end{array}
\end{equation}
The second term in the Eq.~(\ref{eq_17}) is of the order of
${k}^{2}{(d/{L_{H_{0}}})}^{4}\ll{k}^{2}$ and leads to
renormalization of the "effective mass" in a quadratic spectrum
$\varepsilon ({\bf k})\propto {k}^{2}$. This change in the
``effective mass'' does not affect the zero temperature divergence
of the weak-localization correction to the conductance and further
will be neglected. Thus, the resulting spectrum has the form

\begin{equation}\label{res_spec}
\varepsilon({\bf k})={\bf
k}^{2}+\frac{4e^{2}}{{\hbar}^{2}c^{2}}\sum \limits _{{\bf
b}_{n}\not= 0} \frac{{|H_{n}|}^{2}}{{\bf b}_{n}^{2}}.
\end{equation}
Note that the second term in Eq.~(\ref{res_spec}) makes an
important contribution to the dephasing time. The effective
dephasing time has the form

\begin{equation}\label{eq_20}
\frac{1}{\tau _{B} } =\frac{1}{\tau _{\varphi } }
+\frac{4e^{2}D}{{\hbar}^{2} c^{2}}\sum \limits _{{\bf b}_{n}\not=
0} \frac{{|H_{n}|}^{2}}{{\bf b}_{n}^{2}} .
\end{equation}

From the expression (\ref{eq_20}) one can see that for arbitrarily
small magnetic field the low temperature divergence of quantum
correction to the conductance is cut off by the finite effective
dephasing time $\tau _{B}$. The corresponding expression for
quantum correction to the conductance reads:
\begin{equation}\label{eq_19}
\left\langle \Delta g\right\rangle \approx \frac{e^{2} }{2\pi ^{2}
\, \hbar } \ln \left[\frac{\tau }{\tau _{\varphi } } +
\frac{4e^{2}D\tau}{{\hbar}^{2} c^{2}}\sum \limits _{{\bf
b}_{n}\not= 0} \frac{{|H_{n}|}^{2}}{{\bf b}_{n}^{2}} \right].
\end{equation}
The logarithmic term dominates in the weak-localization correction
which allows us to consider the contribution of magnetic field
only in the argument of logarithmic function, and we neglect small
additional corrections to the expression (\ref{eq_19}), which are
also caused by the magnetic field.

In the limit of zero temperature the value $\tau _{B}^{-1} $ is
proportional to the square of amplitude of magnetic field in a
sharp contrast to the case of homogeneous field where $t_{H}^{-1}
\sim H$. This analytical result is in agreement with the
qualitative estimate obtained in Ref.~\onlinecite{wang}.

Now we proceed with the analysis of the case of periodic magnetic
field with nonzero but small spatial average (i.e. $H\not= 0$ and
$H S_{0}\ll \Phi_{0}$, $S_{0}$ is the area of the unit cell
defined by the basic vectors ${\bf a}_{1}$ and ${\bf a}_{2}$) at
low temperatures when $L_{\varphi }\gg d$. The vector potential is
given by

\begin{equation}
\label{A_nonzero} {\bf A}\left({\bf r}\right) =
\frac{1}{2}\left[{\bf H},{\bf r}\right]+\sum \limits_{{\bf b}_{n}
\not= 0} {\bf A}_{n} e^{i {\bf b}_{n} {\bf r}},
\end{equation}
where ${\bf H}=H{\bf z}_{0}$. To obtain the spectrum of the
operator $\hat{H}({\bf r})$ we can first admit that the vector
potential $\ds{\bf A}_{0}({\bf r})=\frac{1}{2}\left[{\bf H},{\bf
r}\right]$ corresponding to the homogeneous component $H$ of
magnetic field is constant on the characteristic scales of a
periodic magnetic field.\cite{peierls} The spectrum of the
operator $\hat{H}({\bf r})$ has the following form (see
(\ref{res_spec})):

\begin{equation}\label{new_spectrum}
\varepsilon({\bf k})={\left({\bf k}-\frac{2e}{\hbar c}{\bf
A}_{0}\right)}^{2}+\frac{4e^{2}}{{\hbar}^{2}c^{2}}\sum \limits
_{{\bf b}_{n}\not= 0} \frac{{|H_{n}|}^{2}}{{\bf b}_{n}^{2}}.
\end{equation}
Proceeding with the analysis in the momentum space one needs to
restore the commutation relations for the components of
quasi-momentum ${\bf k}$ and the components of the radius vector
operator $\hat{\bf r}=i\partial/\partial {\bf k}$. Then the
spectrum (\ref{new_spectrum}) transforms into a new effective
operator, which can be reduced to the harmonic oscillator
hamiltonian. The spectrum of this effective hamiltonian has the
form

\begin{equation}
{\varepsilon}_{m}=\frac{2eH}{\hbar c}
\left(2m+1\right)+\frac{4e^{2}}{{\hbar}^{2}c^{2}}\sum \limits
_{{\bf b}_{n}\not= 0} \frac{{|H_{n}|}^{2}}{{\bf b}_{n}^{2}},
\end{equation}
where $m$ is a nonzero integer number. Finally carrying out the
summation over $m$ in the Eq.~(\ref{eq_15}) we obtain

\begin{equation}
\label{renorm_cond} \begin{array}{c}{\ds \left< \Delta g
\right>=-\frac{e^{2} }{2\pi ^{2} \hbar
}~~~~~~~~~~~~~~~~~~~~~~~~~~~~~~~~~~~~~~~~~~~~~~}\\{\ds
~~~~~~~\times \left\{\psi \left(\frac{1}{2} +\frac{\hbar
c}{4eHD\tau } \right)-\psi \left(\frac{1}{2} +\frac{\hbar
c}{4eHD\tau _{B} } \right)\right\} \ ,}
\end{array}
\end{equation}
where the value ${\tau}_{B}$ is defined by the expression
(\ref{eq_20}). Expanding the expression (\ref{renorm_cond}) in the
limit $HD{\tau}_{B}/{\Phi}_{0}\ll 1$, we obtain

\begin{equation}
\label{eq_newdeltag}
 \left<\Delta g\right>\approx-\frac{e^{2}
}{2\pi ^{2} \, \hbar } \ln \left(\frac{{\tau}_{B} }{\tau } \right)
+\frac{e^{2} }{3\pi ^{2} \hbar } \left(\frac{e{H}D\tau _{B}
}{\hbar c} \right)^{2}.
\end{equation}

The expressions (\ref{renorm_cond}) and (\ref{eq_newdeltag})
formally coincide with the ones for the homogeneous field, but in
a modulated magnetic field  the dephasing time ${\tau}_{B}$ is
determined by the amplitude of modulation.

\section{Strong magnetic fields. Positive magnetoresistance}

\begin{figure*}[hbt!]
\includegraphics[width=0.48\textwidth]{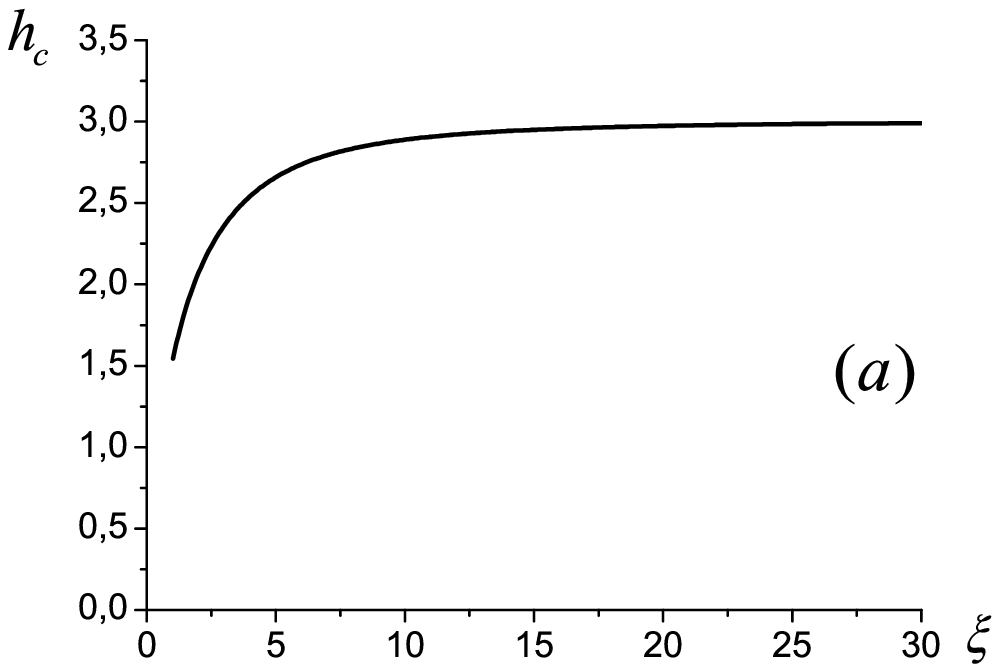}
\includegraphics[width=0.48\textwidth]{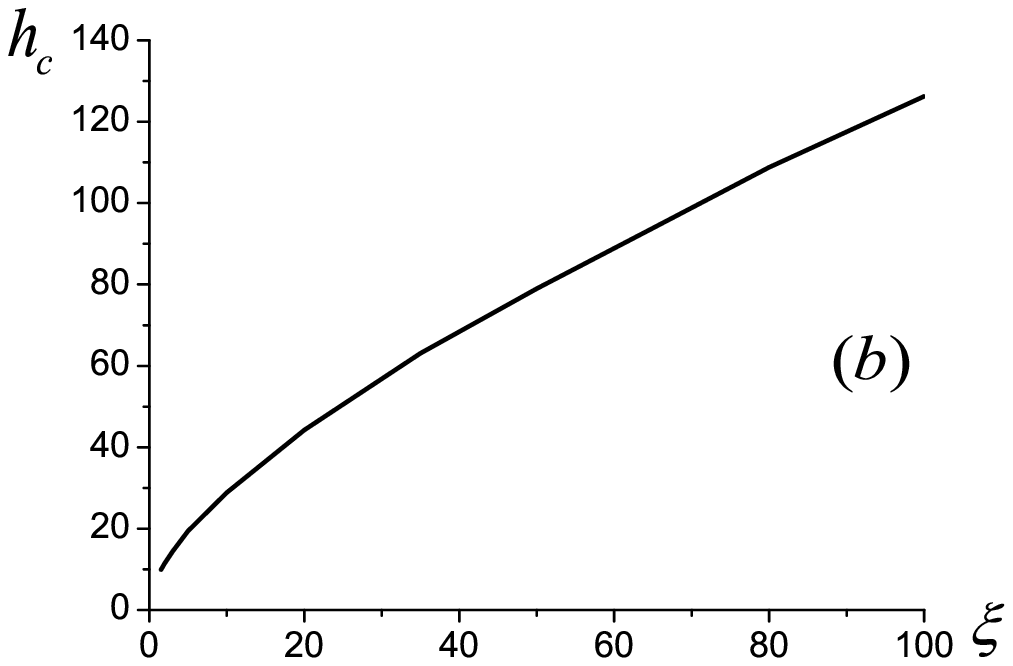}
\caption{The dependencies of the critical magnetic field amplitude
$h_{c}$, which separates the regions of positive and negative
magnetoresistance, vs the parameter $\xi={\tau}_{\varphi}/\tau$:
(a) meander field profile, (b) sinusoidal field profile.}
\label{Fig_6}
\end{figure*}

We now proceed with the consideration of the strong field limit
and focus on the possibility to change the sign of
magnetoresistance of 2DEG in the presence of a modulated magnetic
field. Specifically, we consider a ferromagnetic film/2DEG system
placed in the external magnetic field perpendicular to the 2DEG
plane. Let the ferromagnetic film have a periodic stripe domain
structure. We will assume that the film of 2DEG is thin enough to
consider only the \textit{z} component $H_{z} $ of magnetic field
depending only on $x$ coordinate along the sample surface. We will
denote the  external field as $H$ and the absolute value of the
periodic field of stripe structure by $H_{0} $. Further the
description of the weak-localization correction to the conductance
will be developed on the basis of local approximation. This
approximation is correct when the electron dephasing length is
less than the characteristic scale of inhomogeneous magnetic
field. These conditions mean that $\, d \gg \min \left\{L_{\varphi
}, L_{B} \right\}$ (the gray region in Fig.~\ref{Fig_diagram}).
Note that the effect of positive magnetoresistance can be observed
only in the region of parameters where the local approximation is
applicable. Indeed, the Eqs. (\ref{eq_comb}) and
(\ref{eq_newdeltag}) show that in the opposite limit the second
derivative $\left<\Delta g\left(H\right)\right>$ at $H=0$ is
positive and, as a result, the magnetoresistance is negative.

The effect of positive magnetoresistance strongly depends on the
magnetic field configuration. We assume for simplicity the
thickness of ferromagnetic layer to exceed strongly the period of
stripe structure $d$. In this case for hybrid structures F/2DEG
the spatial configuration of magnetic field in the region of 2DEG
depends mostly on the thickness of the spacer between 2DEG and
ferromagnetic film. If the spacer is much thinner than the period
of stripe structure $d$ then the distribution of the $z$-component
$B_{z}({\bf r})$ of the magnetic field in 2DEG approximately has
the form of meander. In the opposite case, when the spacer
thickness is much larger than the spatial period of the domain
structure, the magnetic field profile is smeared. On a qualitative
level one can describe this limit considering a sinusoidal field
profile.

\subsection{Periodic magnetic field in the form of meander}\label{Sec_3_1}

The periodic magnetic field in the form of meander is the simplest
configuration of magnetic field which reveals the effect of
positive magnetoresistance. The external homogeneous magnetic
field applied to the  system leads to the suppression of weak
localization in the regions where the sign of the external field
coincides with the one of the periodic field component. In
opposite, the external field results in the increase of the
interference corrections in the domains where the sign of these
field components are different. The competition between these two
effects defines the resulting dependence of the averaged
conductance vs an external homogeneous field. If the increase in
the weak-localization correction dominates then the resulting
dependence of averaged conductance vs external homogeneous
magnetic field is decreasing. In this case one can conclude that
2DEG has positive magnetoresistance.

Let us search for the region of parameters, where the effect of
positive magnetoresistance can be observed. Within the local
approximation the averaged conductance
 is defined by
the following expression:
\begin{equation}\label{eq_22}
\ds \left\langle g_{m} (h)\right\rangle =\frac{1}{2}
\left[g\left(h+h_{0} \right)+g\left(h-h_{0} \right)\right] \ ,
\end{equation}
where  $\ds h_{\rm 0}=\frac{4eH_{\rm 0}D\tau _{\varphi } }{\hbar
c} $,
\[\begin{array}{c} {\ds g(H)=g_{D} -\frac{e^{2} }{2\pi
^{2} \hbar }\times} \\ \\ {\ds \left\{\psi \left(\frac{1}{2}
+\frac{\hbar c}{4e\left|H\right|D\tau } \right)-\psi
\left(\frac{1}{2} +\frac{\hbar c}{4e\left|H\right|D\tau _{\varphi
} } \right)\right\} \ ,}\end{array}\] and $g_D$ is the Drude
conductance. Introducing dimensionless variables we obtain:
\begin{equation}\label{eq_21}
g(h)=g_{D} -g_{0} \left\{\psi \left(\frac{1}{2} +\frac{\xi
}{\left|h\right|} \right)-\psi \left(\frac{1}{2}
+\frac{1}{\left|h\right|} \right)\right\},
\end{equation}
where $\ds g_{0} =\frac{e^{2} }{2\pi ^{2} \hbar } $, $\ds\xi
=\frac{\tau _{\varphi } }{\tau } $, $\ds h=\frac{4eHD\tau
_{\varphi } }{\hbar c} $. Expanding the expression (\ref{eq_22})
into the Taylor's series for $h\ll h_{0} $ we find

\[\left\langle g_{m} (h)\right\rangle \approx g\left(h_{0}
\right)+\frac{1}{2} g''\left(h_{0} \right)h^{2} .\] One can see
that the positive magnetoresistance is realized for $h_{0}$, which
satisfies the condition

\begin{equation}\label{eq_23}
g''\left(h_{0} \right)<0.
\end{equation}
This condition is realized when the amplitude of periodic magnetic
field is larger than some critical value $h_{c} $, which depends
on the parameter $\xi$. The dependence $h_{c}(\xi) $ for the
meander configuration of periodic magnetic field is shown in
Fig.~\ref{Fig_6}(a). For $\xi \gg 1$ the boundary of the positive
magnetoresistance region is defined by the condition $h_{c}
\approx 3$.

These conclusions are based on the equation (\ref{eq_3}) and,
thus, are valid only in the diffusive limit. The domain of
applicability of the diffusive approximation is defined by the
condition $L_{B}\gg l$, where $l$ is an elastic scattering length.
In the limit $\lambda_{F}\ll L_{B}\ll l$ (where $\lambda_{F}$ is
the Fermi wave-length) the weak-localization is fully suppressed
and the conductance approaches the Drude value.

\begin{figure*}[hbt!]
\includegraphics[width=0.45\textwidth]{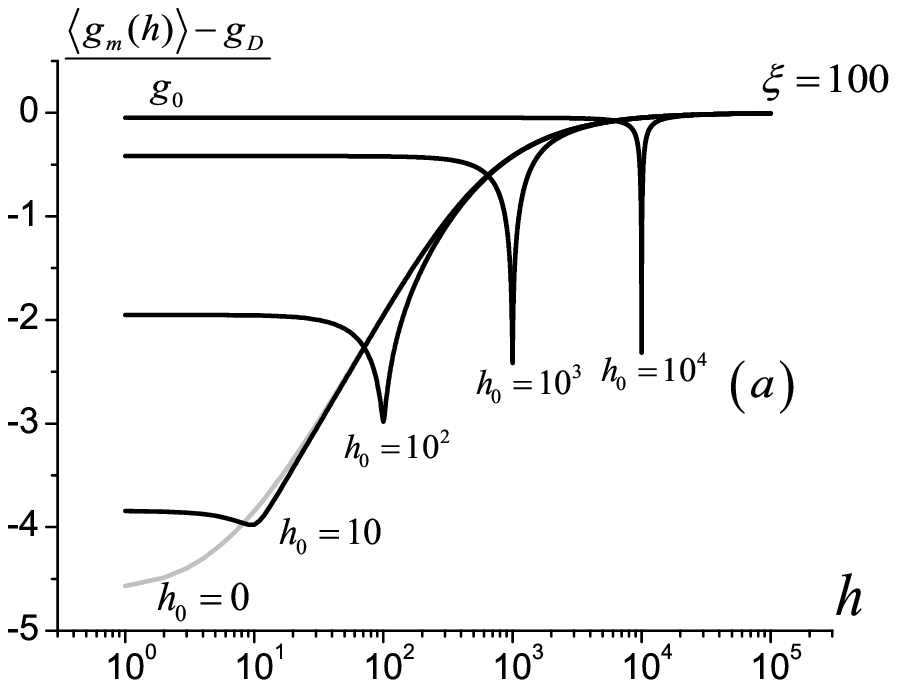}
\includegraphics[width=0.45\textwidth]{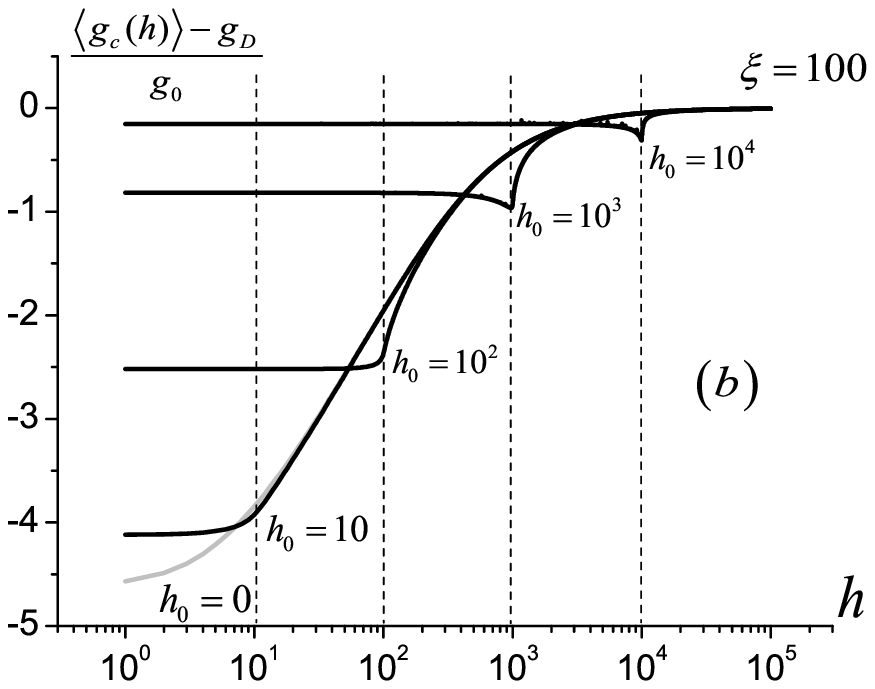}
\caption{The averaged conductance of 2DEG vs the external
homogeneous field at different amplitudes of the periodic field
$h_{0}$ and at $\xi=100$: (a) meander field profile, (b)
sinusoidal field profile. } \label{Fig_7}
\end{figure*}

The dependencies of the averaged conductance of 2DEG vs the
 external homogeneous field at different amplitudes of
periodic field are shown in Fig.~\ref{Fig_7}(a) for $\xi=100$. One
can see that for $h_{0} \gg h_{c} (\xi )$ these dependencies have
the sharp  dips with minima at $h=h_{0}$.

\subsection{Periodic magnetic field in the form of cosine}\label{Sec_3_2}

As a second example we consider the effect of positive
magnetoresistance in the sinusoidal profile of the $z$-component
of the magnetic field
\[H_{z} (x)=H_{0} \cos \left(\pi x/d\right).\]
The expression for the averaged conductance does not depend on the
period of magnetic field $d$ and has the following form:

\begin{equation}\label{eq_24}\begin{array}{c}{\ds
\left\langle g_{c} (h)\right\rangle =g_{D} - g_{0}} \\
{\times \int \limits _{0}^{1}\left\{\psi \left(\frac{1}{2}
+\frac{\xi }{\left|h+h_{0} \cos \left(\pi \rho \right)\right|}
\right)-\psi \left(\frac{1}{2} +\frac{1}{\left|h+h_{0} \cos
\left(\pi \rho \right)\right|} \right)\right\}d\rho.}\end{array}
\end{equation}
Here $\ds h_{0} =\frac{4eH_{0} D\tau _{\varphi } }{\hbar c} $. The
set of dependencies of $\left\langle g_{c} (h)\right\rangle $ for
different magnitudes of amplitude $h_{0} $ is shown in
Fig.~\ref{Fig_7}(b). From the comparison between
Fig.~\ref{Fig_7}(a) and Fig.~\ref{Fig_7}(b) one can see that in
periodic magnetic field with sinusoidal profile the effect of
positive magnetoresistance is weaker than in the case of meander
profile. This is caused by the fact that for the sinusoidal
profile the regions where the external and periodic fields have
opposite directions shrink with the external field increasing.



In Fig.~\ref{Fig_7}(b) the amplitude of the periodic magnetic
field is shown by the vertical dotted line. One can see that for
$h\approx h_{0} $ the behavior of $\left\langle g_{c}
(h)\right\rangle $ changes qualitatively. Even for high amplitude
of the periodic magnetic field $h_{0} $ in the region $h<h_{0} $
the conductance deviates from the Drude value. This is caused by
the incomplete destruction of the interference near the field zero
points even at rather high $h_{0}$ values.


The expression (\ref{eq_24}) allows to find the condition of
 positive magnetoresistance in combined cosinusoidal
and homogeneous magnetic fields. The magnetoresistance peak
appears provided
\[{\left.\frac{{\partial}^{2}{g}_{c}(h)}{\partial{h}^{2}}\right|}_{h=0}<0 ,\] which
gives us the condition
\begin{equation}\label{eq_coscond}
\ds \int \limits _{0}^{1}g''\left(h_{0}\beta
\right)\frac{d\beta}{\sqrt{1-{\beta}^{2}}}<0.
\end{equation}
Here the function $g$ is defined by the expression (\ref{eq_21}).
 The condition (\ref{eq_coscond}) is
satisfied when the periodic magnetic field amplitude $h_{0}$ is
larger than the critical value $h_{c}\left(\xi\right)$. The
dependence $h_{c}\left(\xi\right)$ for sinusoidal profile of
periodic magnetic field is shown in Fig.~\ref{Fig_6}(b). One can
observe a clear difference between two model profiles: contrary to
the meander case the critical field diverges at large $\xi$
values.

\section{Summary}\label{sum}

To sum up, we have investigated the influence of inhomogeneous
magnetic fields on the weak localization phenomenon in 2DEG
systems. In the low field limit we have carried out  a
perturbative analysis of the conductance behavior at high
temperatures and developed an analytical procedure to find a
renormalization of the dephasing rate at low temperatures. In the
high field limit we have justified the validity of the local
approximation and have used this approach to calculate the
averaged conductance for particular model field profiles. It is
found that the systems with modulated magnetic field profiles
provide a possibility to observe the effect of
 positive magnetoresistance.
We have showed that the positive magnetoresistance in
ferromagnetic film/2DEG systems can be observed experimentally
provided the amplitude of the field modulation exceeds a certain
critical value depending on the system parameters.

Finally, we consider some estimates for existing experimental
systems. We take here the 2DEG system in the GaAs/AlGaAs
heterostructure with high electron mobility  as a typical example
(see Ref.~\onlinecite{Typ_2DEG}). This system is characterized by
the following typical parameters: $\hbar c {(4eD\tau)}^{-1}=0.7
Oe$ and $\hbar c{(4eD{\tau}_{\varphi})}^{-1}=0.2 Oe$. The typical
amplitude of the magnetic field induced by the ferromagnetic film
with a domain structure is of the order of ${H}_{0}\sim
10^{2}-10^{3} Oe$ (see Ref.~\onlinecite{alad}). For the meander
distribution of the magnetic field with such amplitude  the height
of the magnetoresistance peak can reach one half of the
weak-localization correction in zero field.

\section*{ACKNOWLEDGEMENTS}\label{Acknow}

This work is supported by the RFBR, RAS under the Federal
Scientific Program ``Quantum physics of condensed matter", the
``Dynasty'' Foundation, the Russian Agency of Education under the
Federal Program ``Scientific and educational personnel of
innovative Russia in 2009--2013".

\end{document}